\documentclass{elsart}

\usepackage{graphicx}

\usepackage{amssymb}

\begin{document}

\begin{frontmatter}

\title{New measurement on photon yields from air and the application 
to the energy estimation of primary cosmic rays}

\author[FUKUI-SCE]{M. Nagano},
\author[FUKUI-ACE]{K. Kobayakawa},
\author[RIKEN]{N. Sakaki} and
\author[FUKUI-APC]{K. Ando}

\address[FUKUI-SCE]{Department of Space Communication Engineering, Fukui
   University of Technology, Fukui, 910-8505 Japan}
\address[FUKUI-ACE]{Department of Architecture and Civil Engineering, 
   Fukui University of Technology, Fukui, 910-8505 Japan}
\address[RIKEN]{RIKEN (The Institute of Physical and Chemical Research), 
   Wako, 351-0198 Japan}
\address[FUKUI-APC]{Department of Applied Physics and
   Chemistry, Fukui University of Technology, Fukui, 910-8505 Japan}

\begin{abstract}
The air fluorescence technique is used to detect ultra-high
energy cosmic rays (UHECR), and to estimate their energy.  Of
fundamental importance is the photon yield due to excitation by
electrons, in air of various densities and temperatures.  After
our previous report, the experiment has been continued using a
\nuc{90}{Sr} $\beta$ source to study the pressure dependence of
photon yields for radiation in nitrogen and dry air.  The photon
yields in 15 wave bands between 300 nm and 430 nm have been
determined.  The total photon yield between 300 nm and 406 nm
(used in most experiments) in air excited by a 0.85 MeV electron
is 3.81$\pm$0.13 ($\pm$13 \% systematics) photons per meter at
1013 hPa and 20 $^{\circ}$C.  The air density and temperature
dependencies of 15 wave bands are given for application to UHECR
observations.
\end{abstract}

\end{frontmatter}

\section{Introduction}
\label{sec:intro}
In order to detect ultrahigh-energy cosmic rays (UHECR),
atmospheric fluorescence light from the trajectory of the
extensive air shower may be measured by mirror-photosensor
systems (the fluorescence technique).  In this type of experiment,
the photon yield from electrons exciting air of various densities
and temperatures is the most fundamental information for
estimating the primary energy of UHECR.  We have reported in our
previous paper \cite{nag03} 
that the photon yield between 300 nm
and 406 nm in air excited by an electron with a mean kinetic
energy of 0.85 MeV (\nuc{90}{Sr} $\beta$) is
3.73$\pm$0.15($\pm$14\% systematics) photons per meter at 1000
hPa and 20$^{\circ}$C.  We measured it with six narrow band
interference filters, whose central wave lengths were 314.7,
337.7, 356.3, 380.9, 391.9 and 400.9 nm.  The bandwidth of each filter
at 50\% of the peak transmission was about 10 nm.  We estimated
the photons in unmeasured wave band to be
8.8\% with help from the values reported by Bunner \cite{bun64}.  We
have also shown that the photon yields in the 337 and 358nm bands are
proportional to d$E$/d$x$, when we compare our measurements with the
photon yields measured at 300$\sim$1000MeV by Kakimoto et
al. \cite{ueno96,kak96}.

We have made the new measurements with additional 9 filters
and have improved the photon yields reported before.  By
comparing the measurements with filters of overlapping
wavelength, the contamination of bands in the tail of each filter
is estimated and corrected.  In this report we provide the
photon yields between 300nm and 430nm from our own measurements.
The air density and temperature dependences of each wave band, and
the average values in the radiative transition from the level
$v'$ to the level $v''$ in the first negative 1N(0, $v''$), and
in the second positive 2P(0, $v''$), 2P(1, $v''$) and 2P(2,
$v''$) systems, are given for application to UHECR observations.

In our previous report \cite{nag03}, we have shown the pressure
dependence of the relaxation rate of the excited level of N$_2$
and N$_2^+$ in nitrogen gas and in air.  The radiative lifetimes
$\tau_0$ have been obtained after determining the
 reference pressure $p'$ from the pressure dependence of 
the photon yields (see Eqs.(5) and (10) in \cite{nag03}).  The 
definitions of $\tau_0$ and $p'$ will be given in Section 3.  Since there
have been many experiments (e.g. \cite{hes68}, \cite{dot73})
before the 1970s that measure $\tau_0$ with accuracy much better
than ours, we don't include our present measurements of $\tau_0$
in this article.

\section{Experiment}
\label{subsec:exp_arrange}

\begin{figure} [thb]
\centerline{\includegraphics[height=8.3cm]{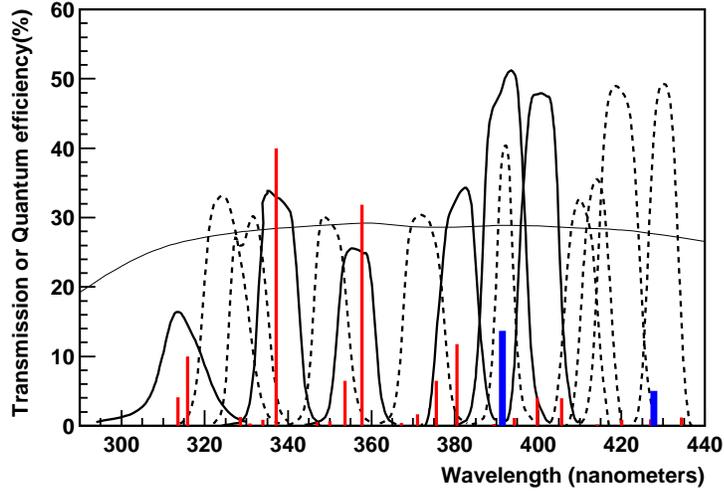}} 
\caption{Typical example of quantum efficiency of PMT used (a thin solid
curve) and transmission coefficients of interference filters used in
the present experiment (dashed curves).  
Thick solid curves are those used in the previous report \cite{nag03}.
The relative intensities of fluorescence
bands at 1000 hPa from the nitrogen molecule (thin vertical lines) and ion
(broad vertical lines) in air from Bunner \cite{bun64} are shown.}
\label{filter}
\end{figure}

We chose a photon counting and thin target technique to measure the
pressure dependence of photon yields (the number of photons produced
by electrons per meter of travel) from nitrogen and air excited by
electrons, following the method employed by Kakimoto et
al. \cite{kak96}. 
Experimental details are described in our previous report\cite{nag03}.

The central values of the new filters used in the present
measurement are 325.0, 330.6, 350.2, 372.5, 410.0, 414.0, 418.5
and 430.0nm, and their bandwidths at half maximum are about 10nm.
A measurement at 392.0nm with a band width of 4.35nm is also
made.  Their transmissions are indicated by dashed curves in
Fig.\ref{filter}.  The solid curves are transmissions of the filters
used in the previous measurement \cite{nag03}.  By subtracting
the contribution of bands in the tail of the transmission curves,
the photon yields from 15 bands between 300 and 430nm are
determined.  Contributions from the 311.7, 313.6, 330.9, 333.9,
346.9, 350.0, 358.2, 367.2, 371.1, 388.4 and 389.4 nm bands can't
be separated and are included in other bands.

The systematic errors of the present experiment are discussed in
\cite{nag03}.  Although the uncertainty from the contamination of
lines in the tails of filter transmissions is reduced from 5\% to
2\%, the total systematic error is only reduced from 13.8\% to
13\%.  This is because the main systematic errors are from
uncertainties in the collection efficiency and quantum efficiency
of the photomultiplier tubes used in the present experiment.

\section{Photon yield}
\label{sec:results}

As described in Bunner \cite{bun64} and 
in our previous report \cite{nag03}, 
the photon yield $\epsilon$ per unit length from gas excited by
an electron is written as a
function of pressure $p$ at a constant temperature $T$ in Kelvin:
\begin{equation}
\epsilon = 
   \frac{p}{R T (h\nu)}\left(\frac{\mbox{d}E}{\mbox{d}x}\right)\
\left(\frac{\Phi^{\circ}}{1 + \frac{p}{p'}} \right)
           = \frac{C p}{1 + \frac{p}{p'}}  \ ,
\label{eq-p} 
\end{equation}
where $R$ is the specific gas constant (N$_2$ : 296.9
m$^2$s$^{-2}$K$^{-1}$ and Air : 287.1 m$^2$s$^{-2}$K$^{-1}$)
, $\displaystyle
\frac{\mbox{d}E}{\mbox{d}x}$ 
is the energy loss in eV kg$^{-1}$ m$^{2}$, and $h\nu$ 
is the photon energy (eV).  $\Phi^{\circ}$
 corresponds to the fluorescence efficiency
in the absence of collisional quenching.
$p'$ is the reference pressure where the lifetime of
collisional de-excitation is equal to the combined lifetime, $\tau_0$, 
 of the excited state
for decay to any lower state, and of internal quenching.

The fluorescence efficiency for the $i$th band at pressure $p$,
 $\Phi_i(p)$, is expressed by
\begin{eqnarray}
\frac{1}{\Phi_i(p)}=\frac{1}{\Phi_i^o}\left(1+\frac{p}{p'_i}\right) \ ,
\label{eff-p}
\end{eqnarray}
where $\Phi_i^o$ is $\Phi^o$ of the $i$th band.

The $p'$  for nitrogen is written in terms of $\tau_o$ and 
the cross-section for nitrogen-nitrogen
collisional de-excitation $\sigma_{nn}$ as
\begin{equation}
   \frac{1}{p'} = \frac {4\tau_o} {\sqrt{\pi M_n kT}}\sigma_{nn} \ ,
\label{ref-pres-n2}
\end{equation}
where $M_n$ is the N$_2$ molecular mass and $k$ is 
Boltzmann's constant.
$p'$ for air is related to $\sigma_{nn}$ and the cross-section for 
nitrogen-oxygen collisional de-excitation $\sigma_{no}$ as 
\begin{equation}
\frac{1}{p'} = \frac{4 \tau_{\circ}}{\sqrt{\pi M_n k T}}
\left(f_n \sigma_{nn} + f_o \sigma_{no}\sqrt{\frac{M_n + M_o}{2M_o}}
\right) = \frac{D}{\sqrt{T}} \ ,
\label{ref-pres-air}
\end{equation}
where  $M_o$ is the mass of oxygen
molecule, and $f_n$ and $f_o$ are the fractions of nitrogen
(0.79) and oxygen (0.21) in air, respectively. 

It should be noted that since $p'$  is
proportional to $\sqrt{T}$, then $p'$ at $T$,  $p'_{T}$,  may be
expressed by 
\begin{equation}
   p'_{T} = p'_{20}\sqrt{\frac {T}{293}} \ ,
\label{p-temp}
\end{equation}
where $p'_{20}$ is $p'$ at 20$^{\circ}$C.

\section{Analysis}
\label{sec:analysis}
\subsection{Derivation of photon yields}
\label{subsec:photon_yield}
The photon yield
per unit length per electron $\epsilon$ is determined as the number of
signal counts $N$ divided by the product of the following: the total
number of electrons $I$, the length of the fluorescence portion $a$,
the solid angle of the PMT $\Omega$, the quartz window transmission
$\eta$, the filter transmission $f$, and the quantum efficiency QE
and the collection efficiency CE of the PMT.

\begin{equation}
 \epsilon = \frac{N}{I \times a \times \Omega \times
  \eta  \times f \times \mathrm{QE} \times \mathrm{CE}} \ .
\label{yield}
\end{equation}
In the following analysis we use a value of $f$ appropriate for
the main nitrogen emission band in each filter band pass. The
number of photons from another band in a given filter is
estimated from measurements in two adjacent filters and is
subtracted from the observed photons.
$I$ is about $(4\sim5)\times10^8$ from about
80 hours in each run.
Finally, $\epsilon$ from a run in a vacuum is
subtracted from each  $\epsilon$ determined above and the
corrected $\epsilon$  is determined.

\subsection{Two components analysis method}
\label{subsec:two_comp}
In order to separate the photons from the 1N band system (427.8 nm) and 
the 2P band system (427.0 nm), 
a ``two components'' analysis has been performed in one filter band.
The method is described in detail in our previous report
\cite{nag03}.
Briefly, we have fitted the observed
pressure dependence of
 $\epsilon_{\mathrm{obs}}(p)$  with
a superposition of two bands in one filter by the least square(LS) 
method.  In this case
the observed photon yield $\epsilon_{\mathrm{obs}}(p)$ 
is the sum of the
photon yields of the main band $\epsilon_{1}(p)$ and the sub-band
$\epsilon_{2}(p)$, and is written as follows :

\begin{eqnarray}
\epsilon_{\mathrm{obs}}(p) = \epsilon_1(p) + \epsilon_2(p)
             &=& \frac{C_1}{x+a_1}
           + \frac{C_2}{x+a_2} \ \ ,
\label{epsilon-two}
\end{eqnarray}
where $x=\frac{1}{p}$, $a_1=\frac{1}{p'_1}$ and
$a_2=\frac{1}{p'_2}$.  $C_1$ and $p'_1$ are the parameters of the
main band in the filter, and $C_2$ and $p'_2$ are those
of the other band.

 We  determine a set of four parameters
$a_1$, $C_1$, $a_2$ and $C_2$ in Eq.(\ref{epsilon-two}) 
by the LS method.

\section{Results}

\label{sec:results}
\subsection{Nitrogen}
\label{subsec:Nitrogen}
\begin{figure}[thb]
\centerline{\includegraphics[height=16.0cm]{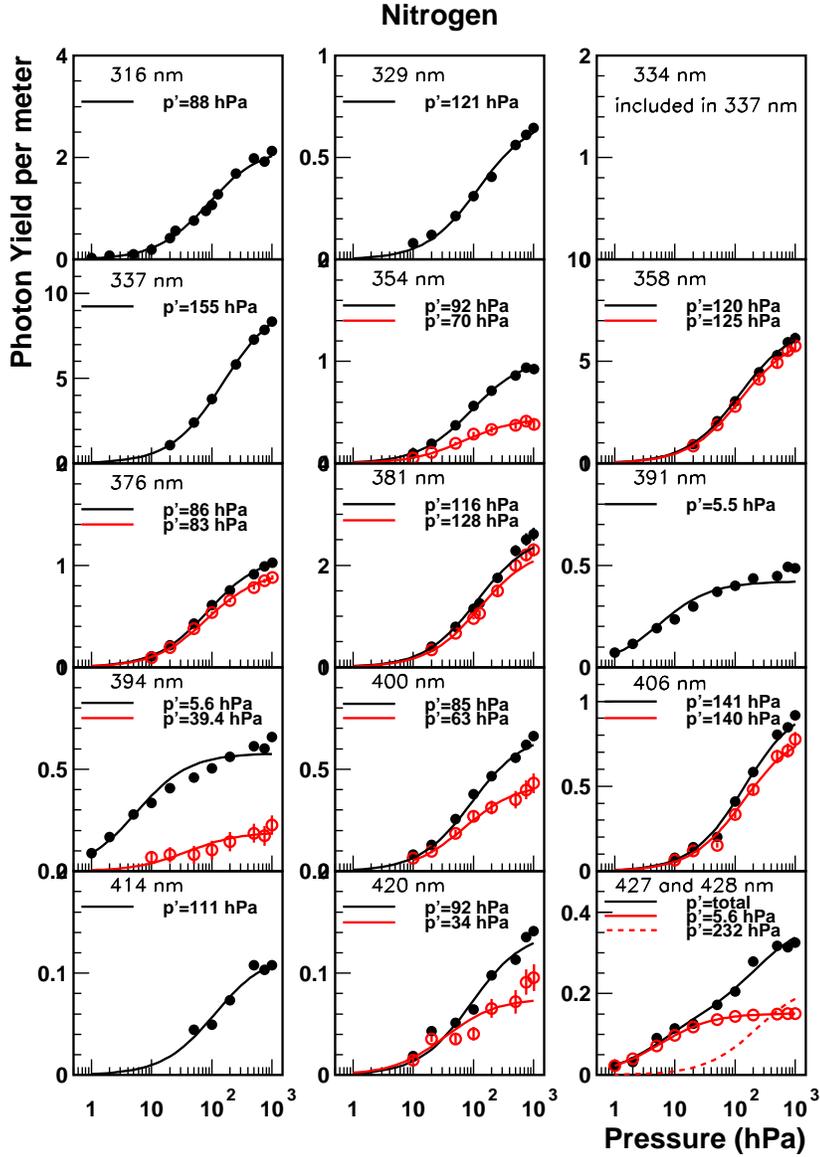}} 
\caption{The pressure dependence of $\epsilon$ in N$_2$ at
20$^{\circ}$C excited by electrons with an average energy of 0.85 MeV
in 14 filters of different central wave length or different wave bands. 
In each figure, the main contributing line in each
wave band is listed. Solid circles are the measured values. 
Open circles are the values after the subtraction of contamination
of other emission bands.
Black solid curves are the best fits of Eq.(\ref{eq-p}) to the experimental
points without subtraction. 
Red ones are the best fits to the corrected points (open circles). 
In the figure labeled 427 and 428 nm, results of a two-components 
analysis are shown.
Open circles with a solid line are from the 1N band system and the dashed line represents
the 2P band system.
}
\label{nitrogen}
\end{figure}

The pressure dependencies of photon yields in nitrogen gas with
fourteen filters are plotted by solid circles in
Fig. \ref{nitrogen}.  In each figure, the main emission band in
each filter band is listed.  The solid curves in the figures are
the best fits of Eq.(1) with $p'$ listed in the figure.  Open
circles are the yields of the main band in the filter after the
subtraction of other bands, which was determined by taking into
account the transmission coefficients of other bands.  The errors
are relatively large due to the uncertainties of these
subtractions.  The values of $p'$ and $C$ for each band are
determined by a fit to Eq. (1) by the LS method as described in
\cite{nag03}.

\begin{table}[htb]
\caption{Summary of the measurements of nitrogen gas.
See text for detail.}

\bigskip
\begin{center}
\catcode`?=\active \def?{\phantom{0}}
\begin{tabular}{|r|c|c|c|c|} \hline
main & \multicolumn{1}{|c|}{$\epsilon$}
& $p'$ & $C$ & \multicolumn{1}{|c|}{$\Phi^{\circ}$}
  \\ \cline{2-5}
$\lambda$(nm)& \multicolumn{1}{|c|}{m$^{-1}$} & hPa & 
 $\times10^{-2}$/(hPa$\cdot$m) & \multicolumn{1}{|c|}{$\times10^{-4}$}
 \\ \hline
316 & ?2.03?$\pm$0.21? & ?88.1?$\pm$?7.5? &
    ?2.51?$\pm$0.14? & ?5.07?$\pm$0.28?   \\ \hline
329 & ?0.622$\pm$0.063 & 121.??$\pm$10.?? & 
    ?0.575$\pm$0.033 & ?1.12?$\pm$0.06?   \\ \hline  
337 & ?8.28?$\pm$0.25? & 155.??$\pm$?4.?? &
    ?6.16?$\pm$0.10? & 11.7??$\pm$0.2??   \\ \hline
354 & ?0.417$\pm$0.044 & ?70.3?$\pm$?6.4? &
    ?0.634$\pm$0.035 & ?1.15?$\pm$0.06?   \\ \hline
358 & ?5.64?$\pm$0.31? & 125.??$\pm$?6.?? &
    ?5.07?$\pm$0.16? & ?9.07?$\pm$0.29?   \\ \hline
376 & ?0.873$\pm$0.059 & ?82.5?$\pm$?4.7? &
    ?1.14?$\pm$0.04? & ?1.95?$\pm$0.07?   \\ \hline
381 & ?2.09?$\pm$0.25? & 128.??$\pm$14.?? &
    ?1.84?$\pm$0.09? & ?3.08?$\pm$0.16?   \\ \hline
391 & ?0.419$\pm$0.049 & ??5.46$\pm$?0.50 &
    ?7.72?$\pm$0.54? & 12.6??$\pm$0.9??   \\ \hline
394 & ?0.185$\pm$0.078 & ?39.4?$\pm$12.5? &
    ?0.49?$\pm$0.13? & ?0.79?$\pm$0.22?    \\ \hline
400 & ?0.399$\pm$0.036 & ?62.9?$\pm$?4.8? &
    ?0.674$\pm$0.033 & ?1.08?$\pm$0.05?    \\ \hline
406 & ?0.73?$\pm$0.15? & 140.??$\pm$25.?? &
    ?0.597$\pm$0.064 & ?0.94?$\pm$0.10?    \\ \hline
414 & ?0.108$\pm$0.029 & 111.??$\pm$24.?? &
    ?0.108$\pm$0.017 & ?0.167$\pm$0.027   \\ \hline
420 & ?0.073$\pm$0.028 & ?34.??$\pm$10.?? &
    ?0.222$\pm$0.050 & ?0.338$\pm$0.076  \\ \hline
427 & ?0.188$\pm$0.113 & 232.??$^{+144.??}_{-?71.??}$ &
    ?0.099$\pm$0.038 & ?0.148$\pm$0.057   \\ \hline
428 & ?0.151$\pm$0.031 & ??5.6?$\pm$?1.1? &
    ?2.72?$\pm$0.24? & ?4.07?$\pm$0.36?   \\ \hline
\multicolumn{1}{|c|}{Sum}  & 21.69?$\pm$0.55? & \multicolumn{3}{|l|}{(300nm$\sim$406nm)} \\ \hline
\multicolumn{1}{|c|}{Sum}  & 22.20?$\pm$0.56? & \multicolumn{3}{|l|}{(300nm$\sim$430nm)}   \\ \hline
\end{tabular}
\end{center}
\label{n2_comp}
\end{table}

Photon yields ($\epsilon$) per meter per electron
of average energy 0.85MeV in nitrogen gas
 at 1013 hPa and 20$^{\circ}$C are determined with $p'$ and
$C$, which
are listed in the third and the fourth column in Table \ref{n2_comp}.
$\Phi^{\circ}$ values are also listed
in the table.  
The total $\epsilon$  between 300 and 406 nm is 21.69$\pm$0.55,
and that between 300 and 435 nm is 22.20$\pm$0.56.

\subsection{Air}
\label{subsec:Air}
\begin{figure}[thb]
\centerline{\includegraphics[height=16.0cm]{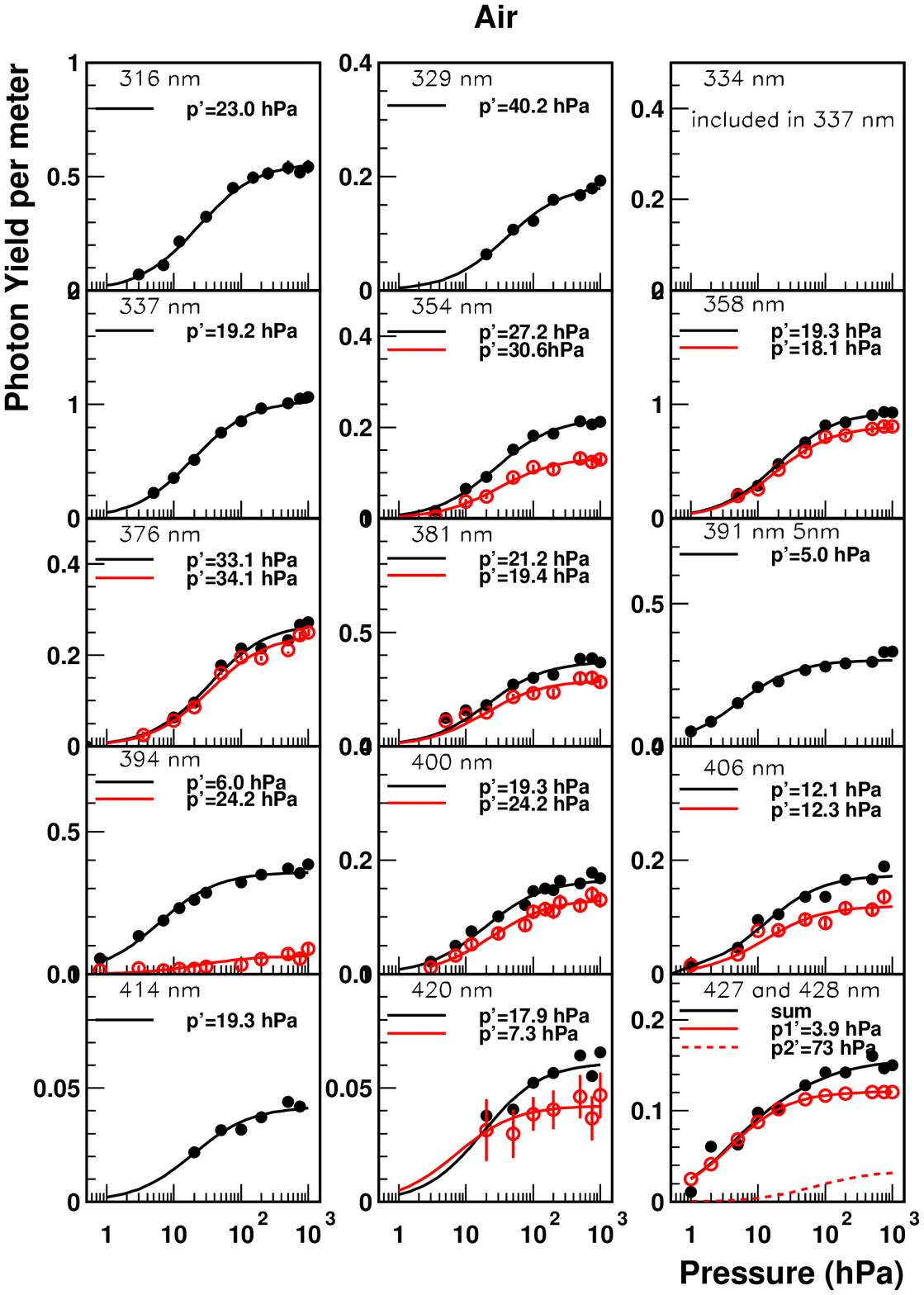}} 
\caption{The pressure dependence of $\epsilon$ in air at
20$^{\circ}$C excited by electrons with an average energy of 0.85 MeV
in fourteen wave bands. In each figure, the main contributing band in each
filter is listed.  Solid circles are the measured values. 
Red open circles are the values after the subtraction of contamination
of other bands described in the text.
 Black solid curves are the best fits of Eq.(\ref{eq-p}) 
 to the experimental points.  Red ones are
the best fits to open circles.
In the plot labeled 427 and 428 nm, 
the results of a two-components analysis are shown.
Open circles with the solid line are from the 1N band system 
and the dashed line represents
the 2P band system.
}
\label{air}
\end{figure}

The pressure dependencies of photon yields in dry air, which is a
mixture of 78.8\% nitrogen gas and 21.1\% oxygen gas, in fourteen
wave bands are indicated by solid circles in Fig. \ref{air}.
Open circles are the yields of the main band in the filter after
the subtraction of other bands, which are estimated from
measurements in neighbouring filter bands, taking into account
the transmission of those lines.

\begin{table}[htb]
\caption{Summary of the measurements of air.}

\bigskip
\begin{center}
\catcode`?=\active \def?{\phantom{0}}
\begin{tabular}{|r|c|c|c|c|} \hline
main & \multicolumn{1}{|c|}{$\epsilon$}
& $p'$ & $C$ & \multicolumn{1}{|c|}{$\Phi^{\circ}$}
 \\ \cline{2-5}
$\lambda$(nm) &  \multicolumn{1}{|c|}{m$^{-1}$} & hPa & 
$\times10^{-2}$/(hPa$\cdot$m) & \multicolumn{1}{|c|}{$\times10^{-4}$} 
\\ \hline
316 & 0.549$\pm$0.057 & ?23.0?$\pm$?1.9?    &
    2.44?$\pm$0.15? & ?4.80?$\pm$0.29?     \\ \hline
329 & 0.180$\pm$0.026 & ?40.2?$\pm$?4.6?    & 
    0.465$\pm$0.042 & ?0.880$\pm$0.080     \\ \hline
337 & 1.021$\pm$0.060 & ?19.2?$\pm$?0.7? &
    5.43?$\pm$0.15? & 10.01?$\pm$0.27?    \\ \hline
354 & 0.130$\pm$0.022 & ?30.6?$\pm$?3.9? &
    0.437$\pm$0.046 & ?0.769$\pm$0.080    \\ \hline
358 & 0.799$\pm$0.080 & ?18.1?$\pm$?1.4? &
    4.50?$\pm$0.28? & ?7.82?$\pm$0.48?    \\ \hline
376 & 0.238$\pm$0.036 & ?34.1?$\pm$?4.1? &
    0.722$\pm$0.068 & ?1.20?$\pm$0.11?   \\ \hline
381 & 0.287$\pm$0.050 & ?19.4?$\pm$?2.6? &
    1.51?$\pm$0.17? & ?2.46?$\pm$0.27?   \\ \hline
391 & 0.302$\pm$0.020 & ??5.02$\pm$?0.26 &
    6.04?$\pm$0.25? & ?9.60?$\pm$0.39?   \\ \hline
394 & 0.063$\pm$0.033 & ?24.2?$\pm$?9.4? &
    0.267$\pm$0.093 & ?0.42?$\pm$0.15?    \\ \hline
400 & 0.129$\pm$0.019 & ?24.2?$\pm$?2.8? &
    0.544$\pm$0.053 & ?0.847$\pm$0.082 \\ \hline
406 & 0.118$\pm$0.019 & ?12.3?$\pm$?1.6? &
    0.972$\pm$0.010 & ?1.49?$\pm$0.15? \\ \hline
414 & 0.041$\pm$0.009 & ?19.3?$\pm$?3.4? &
    0.217$\pm$0.031 & ?0.327$\pm$0.047 \\ \hline
420 & 0.042$\pm$0.015 & ??7.3?$\pm$?1.9?  &
    0.58?$\pm$0.13? & ?0.86?$\pm$0.20?    \\ \hline
427 & 0.032$\pm$0.023 & ?72.??$^{+60.??}_{-23.??}$ &
    0.047$\pm$0.021 & ?0.069$\pm$0.031    \\ \hline
428 & 0.121$\pm$0.022 & ??3.86$\pm$?0.59 &
    3.14?$\pm$0.28? & ?4.57?$\pm$0.41?   \\ \hline
\multicolumn{1}{|c|}{Sum} & 3.81?$\pm$0.13? & \multicolumn{3}{|l|}{(300nm$\sim$406nm)}
   \\ \hline
\multicolumn{1}{|c|}{Sum} & 4.05?$\pm$0.14? & \multicolumn{3}{|l|}{(300nm$\sim$430nm)}
  \\ \hline
\end{tabular}
\end{center}
\label{air_comp}
\end{table}

The values of $\epsilon$ in air at 1013 hPa and 20$^{\circ}$C are
 determined from Eq.(\ref{eq-p}) with the best fitted values of
 $p'$ and $C$.  These together with $\Phi^{\circ}$ values are
 shown in Table \ref{air_comp}.  In estimating $\epsilon$, $C$
 and $\Phi^{\circ}$, the contamination of lines in the tail of
 the transmission curve are subtracted from each other as
 mentioned before.  Therefore some values listed in this table
 are somewhat reduced from the values in Table 9 of \cite{nag03},
 where the measured values in each filter at 1000 hPa were
 listed.  Total photons per meter per electron between 300 and
 406nm is 3.81$\pm$0.13 and between 300 and 428nm is
 4.05$\pm$0.14.

\section{Discussion on experimental results}
\label{sec:discussion}
\subsection{Comparison with our previous report}

\begin{figure}[thb]
\centerline{\includegraphics[height=10.5cm]{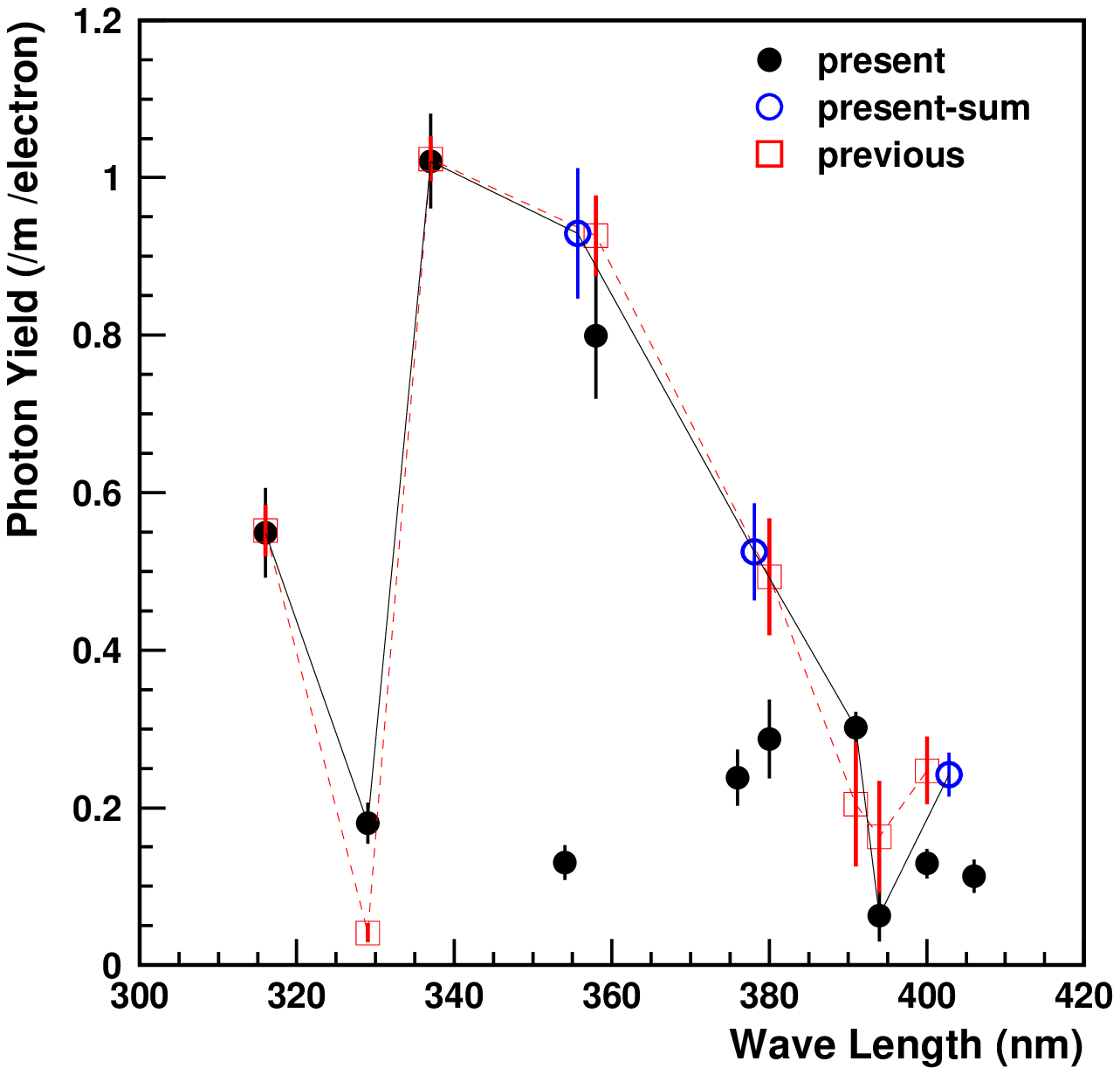}}
\caption{Comparison of the present photon yields per meter per electron
(closed circles) with our previous report (open squares)
\cite{nag03}. Open circles are the sum of two bands which were
not separated in our previous measurement.}
\label{new-pre}
\end{figure}

The present results are obtained after subtracting 
the contamination of other bands in the tail of 
each filter, by comparing the measurements with another
filter of overlapping wavelength.  In this subtraction
we have used the fitted curve to the  experimental
points. The photon yields at 1013 hPa in Table
\ref{n2_comp} and \ref{air_comp} are 
calculated with the best
fitted values of $p'$ and $C$. In contrast, in our previous
report, the measured values at 1000 hPa are
listed and the unmeasured values are estimated
from the table by Bunner \cite{bun64}.

In Fig. \ref{new-pre}, the present values of $\epsilon$ per meter per
electron at 1013 hPa and 20$^{\circ}$C are compared with the
previous results.  The closed circles are the present results and
the open circles are the sum of two bands where they were not
separated in the previous measurement.  The lines are drawn to
compare them easily.  There are slight differences between the
new and old results at 329 nm and 391 nm.  In the previous report
we estimated the value at 329 nm from that listed in Bunner
\cite{bun64}, which is relatively low compared with the present
measurement.  Considering the 391 nm value, in the previous
report we separated the 394nm band from the 391nm band by using a
two components analysis, while in the present experiment the
filter of the narrower bandwidth is used to separate them. Though
the combined yield at 391nm and 394nm is similar, the individual
values are somewhat different as seen around 391nm in
Fig. \ref{new-pre}.
 
\subsection{Density and temperature dependence of the photon yields}
The photon yields per meter by an electron of energy $E$
can be rewritten as a function of the
gas density  $\rho$ (in kg m$^{-3}$)
and the temperature (in Kelvin) to apply to the
atmosphere  \cite{kak96}. 

\begin{eqnarray}
\epsilon &=& \frac{\left(\frac{\d E}{\d x}\right)_{E}}
{\left(\frac{\d E}{\d x}\right)_{0.85\mathrm{MeV}}}
\frac{A\rho}{1+\rho B \sqrt{T}} \ ,
\label{eq-t-den} 
\end{eqnarray}
where
\begin{equation}
 A = \frac{\left(\frac{\d E}{\d x}\right)_{0.85\mathrm{MeV}} \Phi^{\circ}}{h \nu}
\end{equation}
and 
\begin{equation}\
 B = \frac{R_{air}}{p'_{20}}\sqrt{293} \ .
\end{equation}

where $R_{air}$ is the specific gas constant for air (see
equation \ref{eq-p}) .  The values of $A$ and $B$ are calculated
and are listed in Table \ref{AandB}.

\begin{table}[t]
\caption{A and B values in 15 bands}

\bigskip
\begin{center}
\catcode`?=\active \def?{\phantom{0}}
\begin{tabular}{|r|c|c|c|c|} \hline
main & \multicolumn{2}{|c|}{Nitrogen} & \multicolumn{2}{|c|}{Air} \\ 
\cline{2-5}
$\lambda$(nm) & $A$ & \multicolumn{1}{|c|}{$B$} & $A$ & 
\multicolumn{1}{|c|}{$B$} 
 \\ \cline{2-5} 
     &  m$^2$kg$^{-1}$ &   m$^3$kg$^{-1}$K$^{-\frac{1}{2}}$ 
     &  m$^2$kg$^{-1}$ &   m$^3$kg$^{-1}$K$^{-\frac{1}{2}}$   \\ \hline
316 & 21.8?$\pm$1.2? & ?0.577$\pm$0.049 
  & 20.5?$\pm$1.3? & ?2.14$\pm$0.18  \\ \hline 
329 & ?5.00$\pm$0.29 & ?0.419$\pm$0.035  
  & ?3.91$\pm$0.35 & ?1.22$\pm$0.14  \\ \hline
337 & 53.6?$\pm$0.9? & ?0.328$\pm$0.008 
  & 45.6?$\pm$1.2? & ?2.56$\pm$0.10 \\ \hline
354 & ?5.52$\pm$0.31 & ?0.723$\pm$0.066 
  & ?3.68$\pm$0.39 & ?1.60$\pm$0.21 \\ \hline
358 & 44.1?$\pm$1.4? & ?0.407$\pm$0.019 
  & 37.8?$\pm$2.3? & ?2.72$\pm$0.22 \\ \hline
376 & ?9.95$\pm$0.35 & ?0.616$\pm$0.035 
  & ?6.07$\pm$0.57 & ?1.44$\pm$0.17  \\ \hline
381 & 16.0?$\pm$0.8? & ?0.397$\pm$0.042 
  & 12.7?$\pm$1.4? & ?2.53$\pm$0.35  \\ \hline
391 & 67.2?$\pm$4.7? & ?9.31?$\pm$0.86? 
  & 50.8?$\pm$2.1? & ?9.80$\pm$0.51  \\ \hline
394 & ?4.3?$\pm$1.2? & ?1.29?$\pm$0.41?  
  & ?2.25$\pm$0.78 & ?2.03$\pm$0.79  \\ \hline
400 & ?5.87$\pm$0.28 & ?0.808$\pm$0.061  
  & ?4.58$\pm$0.44 & ?2.03$\pm$0.23  \\ \hline
406 & ?5.20$\pm$0.56 & ?0.363$\pm$0.064 
  & ?8.18$\pm$0.82 & ?3.99$\pm$0.52  \\ \hline
414 & ?0.94$\pm$0.15 & ?0.46?$\pm$0.10?  
  & ?1.83$\pm$0.26 & ?2.55$\pm$0.45  \\ \hline
420 & ?1.93$\pm$0.43 & ?1.49?$\pm$0.46?  
  & ?4.9?$\pm$1.1? & ?6.8?$\pm$1.7?  \\ \hline
427 & ?0.86$\pm$0.33 & ?0.22?$\pm$0.10?
  & ?0.40$\pm$0.18 & ?0.68$\pm$0.38  \\ \hline 
428 & 23.7?$\pm$2.1? & ?9.1??$\pm$1.8??  
  & 26.5?$\pm$2.4? & 12.7?$\pm$1.9?  \\ \hline
\end{tabular}
\end{center}
\label{AandB} 
\end{table}

\begin{figure}[thb]
\centerline{\includegraphics[height=10.5cm]{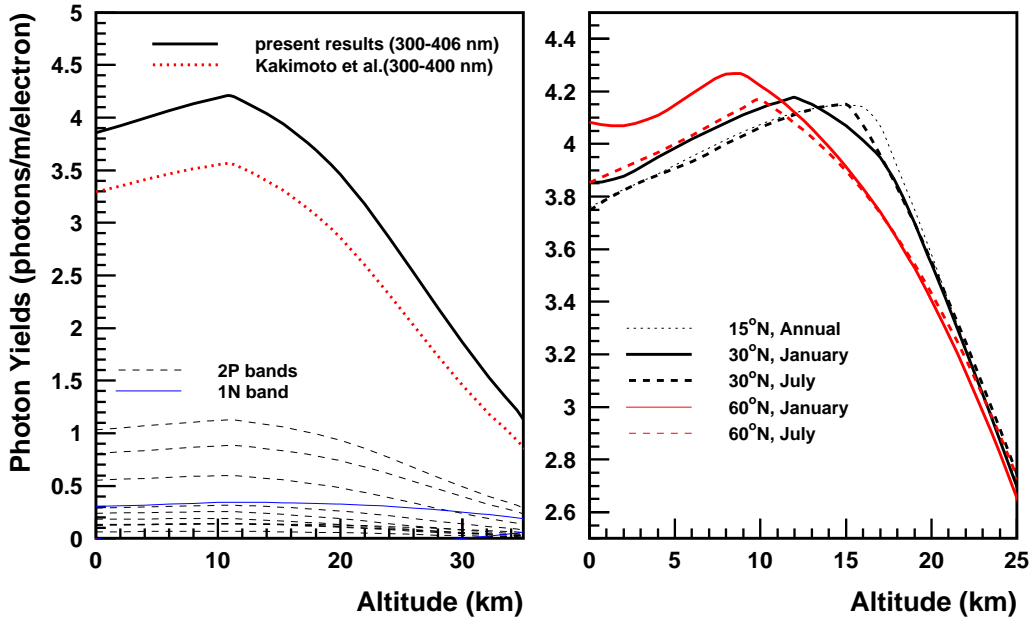}}
\caption{The altitude dependence of photon yields per meter for
a 0.85 MeV electron ( between 300 and 406 nm). In the right- 
and left-hand figures, the US Standard Atmosphere 1976 and
1966 are assumed, respectively.  In the left-hand
figure, thin dashed lines are from the 2P band systems and a thin
line is from the 1N band system. A thick solid line is the
sum of all bands between 300 and 406 nm.  The relation from
Kakimoto et al. is also shown with a dotted line for wavelengths
between 300 and 400 nm.  The photon yields in January and July
are compared in the right-hand figure at latitudes of
30$^{\circ}$N and 60$^{\circ}$N.  }
\label{alt}
\end{figure}

Using the values of $A$ and $B$ in Table \ref{AandB}, the photon
yield of a 0.85 MeV electron is shown in Fig. \ref{alt} as a
function of altitude.  The US Standard Atmosphere 1976
\cite{rika02} is assumed in the left-hand figure.  Thin dashed
lines are from the 2P band system and a thin solid line is from
the 1N band system. A thick line shows the sum from all bands
between 300nm and 406nm.  The altitude dependence from Kakimoto
et al.  between 300 nm and 400 nm is also shown by a thick dotted
line.  Since the proportion of yield from the 406nm band is
3.2\%, there is a difference of about 14\% (300$\sim$400 nm)
between ours and Kakimoto et al. independent of altitude.  This
is because the photon yields of Kakimoto et al.  were
underestimated in their unmeasured bands.

The altitude dependence of photon yields in January and July are
compared in the right-hand figure for latitudes 30$^{\circ}$N and
60$^{\circ}$N, assuming the US Standard Atmosphere 1966
\cite{us66}.  The annual value is also shown for 15$^{\circ}$N.
Monthly mean values of temperature and densities from CIRA 1986
(COSPAR International Reference Atmosphere) lie approximately
between the lines shown for January and July at 30$^{\circ}$N
\cite{vin03}.  It would be necessary to use the several types of
altitude dependence at different latitudes in experiments from
space like EUSO (Extreme Universe Space Observatory)
\cite{euso00}.

\subsection{Average $p'$ values in different radiative systems}
For the radiative transition from the level $v'$ to the level $v''$
in the 1N and 2P systems,
$p'$ values are expected to be the same for a fixed 
 $v'$ irrespective of
$v''$, if $\Phi^{\circ}$ is much smaller than unity.
So in Table \ref{average}, we summarize the
average values of  $p'$,  $B$ and the ratio of photon yields
in nitrogen gas and air at 1013 hPa.
Here weighted averages of four bands (337, 358, 381 and 406nm) for
2P(0, $v''$), five bands (316, 354, 376, 400 and 427 nm) for
2P(1, $v''$), two bands (394 and 420 nm) for
2P(2, $v''$) and two bands (391 and 428 nm) for
1N(0, $v''$) are taken for each parameter.

The present  $p'$ values are larger than those
listed in Bunner \cite{bun64}, except 2P(0,$v''$) in air.
The reason is not clear. In the next sub-section we compare the
fluorescence efficiency at pressures of 100 $\sim$ 1000 hPa
directly with other experiments. 

\begin{table}[htb]

\medskip
\caption{Average $p'$ and $B$ values 
in different radiative systems in nitrogen gas and in air.}

\bigskip
\begin{center}
\catcode`?=\active \def?{\phantom{0}}
\begin{tabular}{|r|c|c|c|c|c|} \hline
transition & gas & \multicolumn{1}{|c|}{$p'$}
& \multicolumn{1}{|c|}{$B$} &  $\epsilon(N_2)$/$\epsilon(air)$ 
& \multicolumn{1}{|c|}{$p'$ \ \ [2]} \\ \cline{2-4}  \cline{6-6} 
  state &   &  \multicolumn{1}{|c|}{hPa at 20$^{\circ}$C} 
&  m$^3$kg$^{-1}$K$^{-\frac{1}{2}}$  &  at 1013 hPa 
&  \multicolumn{1}{|c|}{hPa at 27$^{\circ}$C} \\ \hline
2P(0,$v''$) & N$_2$  &144.7?$\pm$3.1? &  ?0.343$\pm$0.013 &
               7.70$\pm$0.42 & 120??  \\ \cline{2-4} \cline{6-6}
            & air    & ?18.1?$\pm$0.6? &  ?2.72?$\pm$0.09?   
            & & ?20?? \\ \hline
2P(1,$v''$) & N$_2$  & ?74.5?$\pm$2.8? & ?0.628$\pm$0.03 &
               3.43$\pm$0.29 & ?32.6  \\ \cline{2-4} \cline{6-6}
            & air    & ?25.6?$\pm$1.4? & ?1.92?$\pm$0.10?  
            & & ??8.7   \\ \hline
2P(2,$v''$) & N$_2$  & ?36.2?$\pm$8.0? & ?1.38?$\pm$0.26? &
               1.94$\pm$0.81 & ?14.5  \\ \cline{2-4} \cline{6-6}
            & air    & ??7.9?$\pm$1.8? & ?6.2??$\pm$1.4??  
            & & ??6.1   \\ \hline
1N(0,$v''$) & N$_2$  & ??5.48$\pm$0.46 & ?9.27?$\pm$0.25? &
              1.36$\pm$0.16 & ??1.6 \\ \cline{2-4} \cline{6-6}
            & air    & ??4.83$\pm$0.24 & 10.2??$\pm$0.5??  
            & & ??1.6  \\ \hline
\end{tabular}
\end{center}
\label{average} 
\end{table}

\subsection{Comparison with other experiments}
\label{subsec:comp}
The present values of $\Phi_i(p)$ given by Eq.(\ref{eff-p}) in
N$_2$ and air at $p=$800 hPa ($\sim$600 torr) are listed in Table
\ref{summary}, together with those of Davidson and O'Neil
\cite{dav64}, Kakimoto et al.\cite{ueno96,kak96} and Mitchell
\cite{mit70}.  The values of Davidson and O'Neil are on average
2.6 times larger than ours for nitrogen, but 0.88 of ours for
air. The agreement of the present results with Kakimoto et al.,
whose experimental conditions are similar to the present ones, is
rather good both in nitrogen and in air.  The efficiencies of
several bands calculated from the experiment of Mitchell using
X-ray photons from 0.9 to 8.0 keV \cite{mit70} are also listed.
Though his excitation method and the target thickness are quite
different from ours, it should be noted that the efficiencies at
800 hPa for air are in fairly good agreement with each other.

\begin{table}[htb]
\medskip
\caption{Comparison of fluorescent efficiencies $\Phi_i(p)$ for N$_2$
and air at $p$=800 hPa with those of D\&O:Davidson and O'Neil \cite{dav64},
K:Kakimoto et al. cited from \cite{ueno96},\cite{kak96}
and M:Mitchell \cite{mit70}.}

\bigskip
\begin{center}
\catcode`?=\active \def?{\phantom{0}}
\begin{tabular}{|r|c|c|c|c|c|c|c|c|} \hline
wave   & \multicolumn{4}{|c|}{Nitrogen}
       & \multicolumn{4}{|c|}{Air}  \\ \cline{2-9}
length & \multicolumn{1}{|c|}{Present} & \multicolumn{1}{|c|}{D\&O} 
& K & M
       & \multicolumn{1}{|c|}{Present} & \multicolumn{1}{|c|}{D\&O} &
 K & M  \\ \cline{2-9}
 nm    &  \multicolumn{4}{|c|}{$\times10^{-4}$} & 
 \multicolumn{4}{|c|}{$\times10^{-5}$}  \\ \hline
337 & 1.90?$\pm$0.06? & 5.20 & 1.87 & 2.45  & 2.34?$\pm$0.11? & 2.10 & 2.1 & 3.38 \\ \hline 
354 & 0.093$\pm$0.010 & 0.47 &      &       & 0.284$\pm$0.047 & 0.32 &    & \\ \hline
358 & 1.23?$\pm$0.07? & 3.70 & 1.35 &       & 1.73?$\pm$0.17? & 1.50 & 2.2 &   \\ \hline 
376 & 0.182$\pm$0.012 & 0.37 &      &       & 0.489$\pm$0.074 & 0.30 &  &  \\ \hline 
381 & 0.425$\pm$0.050 & 1.40 &     & 0.585 & 0.58?$\pm$0.10? & 0.52 &  & 0.79 \\ \hline 
391 & 0.086$\pm$0.010 & 0.10 &      & 0.114 & 0.598$\pm$0.040 & 0.70 & 0.84 & 0.64 \\ \hline 
394 & 0.037$\pm$0.015 &      &      &       & 0.123$\pm$0.065 & 0.05 &  &  \\ \hline 
400 & 0.079$\pm$0.007 & 0.20 &      &       & 0.249$\pm$0.037 & 0.18 &  &  \\ \hline 
406 & 0.140$\pm$0.029 & 0.43 &      & 0.180 & 0.226$\pm$0.017 & 0.18 &  & 0.227 \\ \hline 
428 & 0.028$\pm$0.006 & 0.11 &      &       & 0.220$\pm$0.039 & 0.27 &  &  \\ \hline 
\end{tabular}
\end{center}
\label{summary} 
\end{table}

\begin{figure}
\centerline{\includegraphics[height=15.0cm]{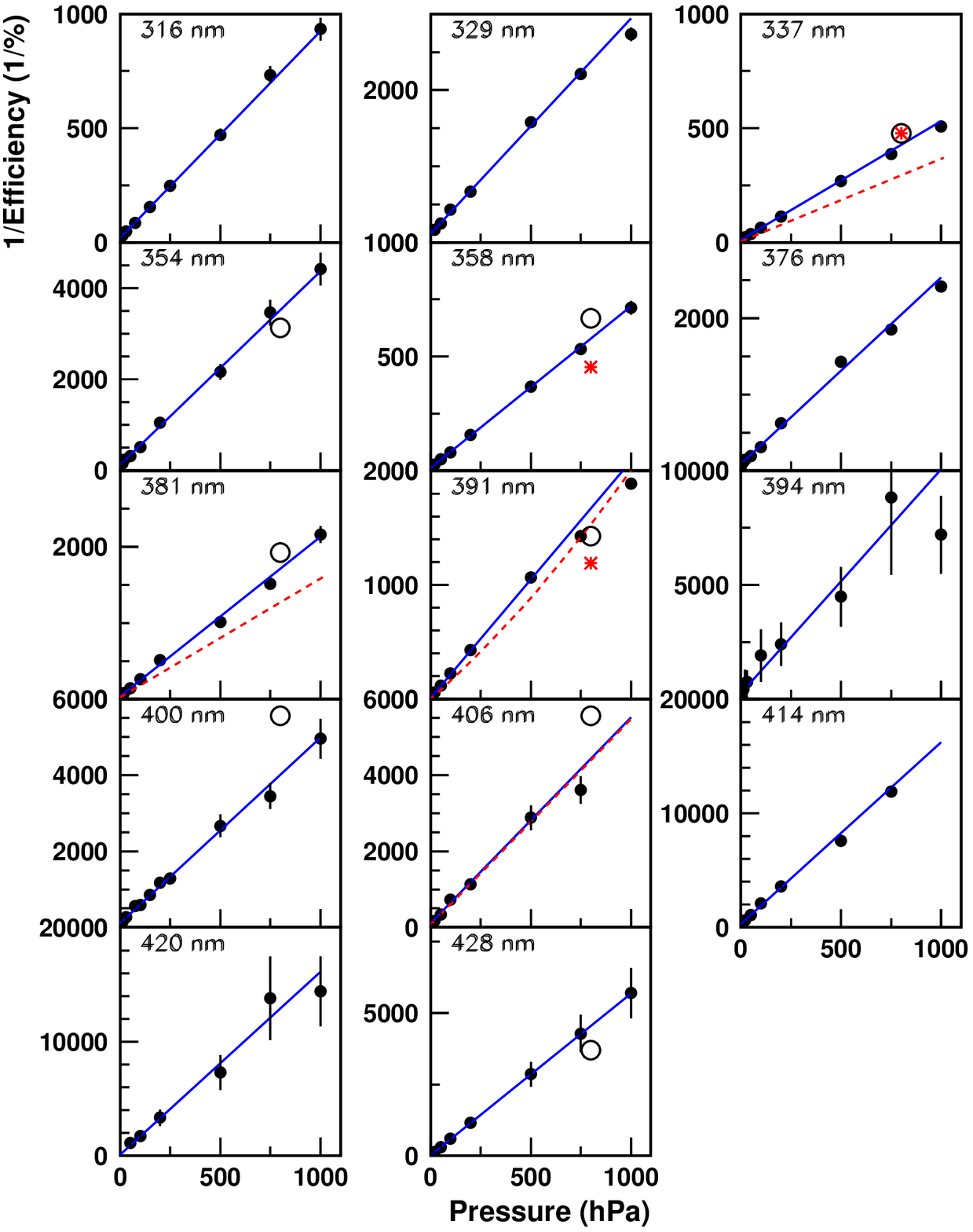}}
\caption{The reciprocal of the efficiency $\frac{1}{\Phi(p)}$ in air 
is plotted
as a function of pressure.  The best-fit functions are shown by
 solid lines.  The relations of Mitchell \cite{mit70} are shown as
dashed lines in some bands. Open circles are from   
 Davidson and O'Neil \cite{dav64}. Asterisks are from Kakimoto et al.
\cite{kak96}
}
\label{efficiency}
\end{figure}

The efficiency $\Phi_i^o$ has been determined from the intercept
at zero pressure of Eq.(\ref{eff-p}) in past measurements at low
pressures \cite{hir70}.  $\frac{1}{\Phi_i(p)}$ for air in
$\frac{1}{\%}$ is plotted as a function of pressure in
Fig. \ref{efficiency}.  The results of LS fitting to
Eq.(\ref{eff-p}) are shown by solid lines.  The determined
$\Phi_i^o$ from the intercept at zero pressure and $p'$ from the
slope are in good agreement with the values listed in Table
\ref{air_comp}.  The dashed lines are from Mitchell \cite{mit70}.
The agreement is good for the 391 and 406 nm bands, but not good
at 337 nm and 381 nm.  Open circles are from Davidson and O'Neil
\cite{dav64} and the asterisks are from Kakimoto et
al. \cite{kak96}.  Their agreement with the present results for
the 337, 354, 358, 381, 391 and 428 nm bands are good, while
there are slight differences for the 400 and 406 nm bands.

As shown in Fig. \ref{efficiency}, the present results are in
fairly good agreement with Davidson and O'Neil \cite{dav64} and
Mitchell \cite{mit70}.  The values of $p'$ and $\Phi^o$ for the 391
nm band from Mitchell are in good agreement with Hirsh et
al. \cite{hir70} measured at low pressure.  According to
Mitchell, for the 391 nm band there is a significant deviation
from Eq. (\ref{eff-p}) at pressures higher than 100 torr (133
hPa).  This deviation results from a three-body deactivation
process and can be expressed as
\begin{eqnarray}
\frac{1}{\Phi(p)}=\frac{1}{\Phi^o}(1+ K p + D p^2),
\label{eff-p-391}
\end{eqnarray}
where $\Phi^o$=0.53\%, $K$=1.08torr$^{-1}$ (0.810hPa$^{-1}$) and
$D$=4.4$\times10^{-4}$torr$^{-2}$ (2.48$\times10^{-4}$hPa$^{-2}$)
for the 391 nm band, which is shown by a dashed curve in the
figure.
Therefore we may conclude that our efficiencies 
at high pressures coincide with other data
 within experimental error, even if
the present $p'$ is different from that determined
at low pressure by Hirsh et al.

\begin{figure}[thb]
\centerline{\includegraphics[height=10.0cm]{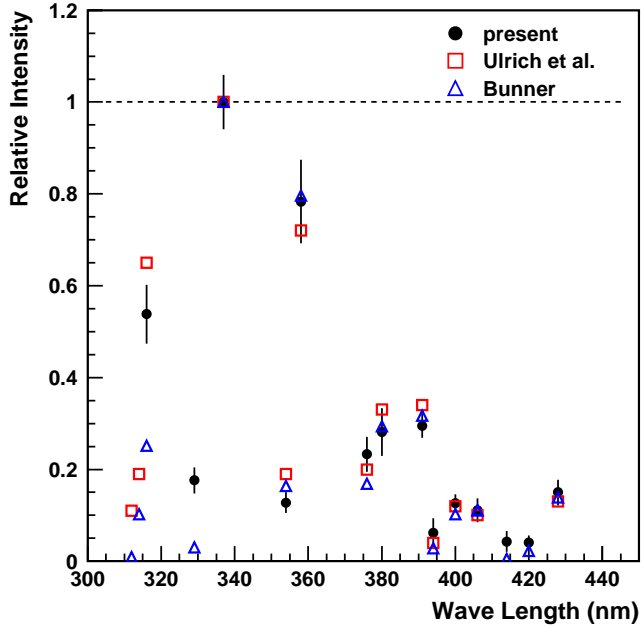}}
\caption{Relative photon intensities in air at 1013 hPa in each band
 between 300nm and 430 nm. The intensities are normalized to 
1.0 at 337 nm. 
}
\label{relative}
\end{figure}

The relative intensity of each band in air is plotted
in Fig. \ref{relative} with those from Bunner \cite{bun64}
and Ulrich et al.\cite{ulr03}.  The incident energies of 
electrons are quite different (present:0.85MeV, 
Ulrich et al.:15keV 
Bunner:mainly 50keV), but 
the agreement may be within the experimental error of
each experiment.

We may, therefore, surely conclude that the present results can
be applied to cosmic ray experiments, where photon yields at high
pressures above 100 hPa are important.

\section{Remarks to the energy determination of UHECR}
\subsection{Effects to the UHECR experiments so far}

In order to estimate the effects of the present results to the
energy determination of the UHECR experiments so far made, let us
compare the observed number of photons from the extensive air
showers using the present photon yields with those used by the
HiRes experiment \cite{daw02}.  The HiRes experiment currently
uses a combination of results from Kakimoto et al. \cite{kak96}
and Bunner \cite{bun64}.

The conditions for the present calculation are similar
to a previous one \cite{sak03} and 
are summarized below.

\begin{itemize}
 \item Air shower simulation code used: CORSIKA 6.020\ \cite{corsika,corsika_upgrade} with QGSJET model \cite{kal97}.
  \begin{itemize}
       \item Primary particle=proton
       \item Energy $E$=$10^{19}$eV and $10^{20}$eV 
       \item Zenith angle  $\theta=0$ and 60$^\circ$ 
       (In the case of $\theta=60^\circ$, the shower axis is on the plane
       perpendicular to the line of sight.)
       \item In each combination of above parameters,
         30 events are simulated to get the average longitudinal 
         development curve.
   \end{itemize}
 \item The number of emitted photons is calculated from the total energy
       deposit of shower particles, $\Sigma$d$E$/d$x$, in each 
       step for sampling of longitudinal development, 
    using Eq. (\ref{eq-t-den}).  The energy release provided by CORSIKA
   was used for this calculation, in which the contribution by particles
   below simulation energy threshold is carefully taken into account
   \cite{ris04}.
 \item The observation height is 0 m a.s.l.
 \item Emitted photons in each step width of depth are attenuated 
by Rayleigh scattering with the following transmission factor
\begin{equation}
 T_R=\exp\left[-\frac{|x_1-x_2|}{X_R}\left(\frac{400\mathrm{[nm]}}{\lambda}\right)^4\right],
\end{equation}
       where $X_R$=2974 g/cm$^2$, and $x_1$ and $x_2$ 
are the slant depths of two points.
 \item Photons are attenuated by Mie scattering with
\begin{equation}
 T_M=\exp\left(\frac{H_M}{L_M \cos \theta} \left[\exp\left(-\frac{h_1}{H_M}\right)-\exp\left(-\frac{h_2}{H_M}\right)\right]\frac{400\mathrm{[nm]}}{\lambda}\right),
\end{equation}
       where scale height $H_M=1.2$km and the horizontal attenuation 
    length $L_M=25$km. $h_1$ and $h_2$ are the heights of the 
     emission point and the detection point of the light, respectively.
\item Total number of observed photons is calculated by adding photons
   from each step of depth.
\item US standard atmosphere 1976\cite{rika02} is used for the
 altitude dependences of density and temperature.
\item  Wavelength dependences of the HiRes filter transmission  and quantum
       efficiency (QE) of the HiRes Photomultiplier tube 
       (PMT) are taken into account
       in some cases.
\end{itemize}

\begin{figure}[thb]
\centerline{\includegraphics[width=14.0cm]{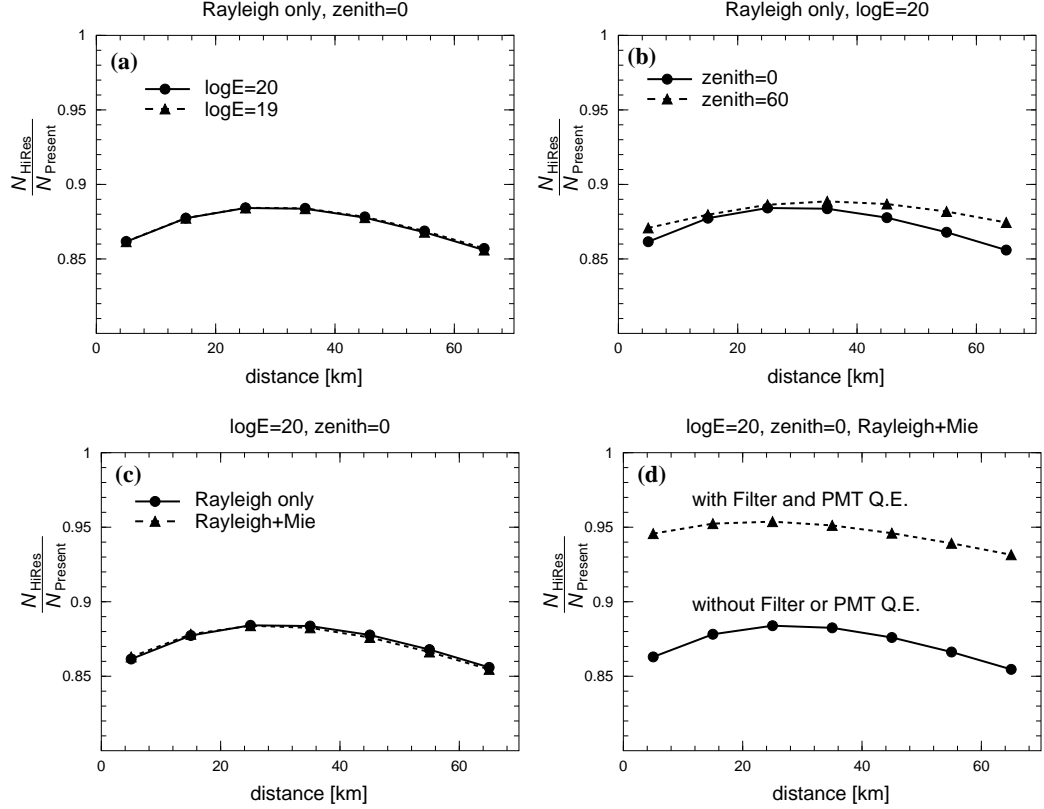}}
\caption{Comparison of observed total photon number in various
 conditions. $R_N=\frac{N_{HiRes}}{N_{present}}$ is plotted as
a function of distance. 
(a) The effect of the primary energy for
 the showers with $\theta=0^\circ$.
 Circles with a solid line are for $\log E(\mathrm{eV})=20$
 and triangles with a dashed line for $\log E(\mathrm{eV})=19$.
 Only Rayleigh scattering is taken
 into account.
 (b) The effect of zenith angle for $\log E=20$. Only Rayleigh
 scattering is taken into account. A solid line is for $0^\circ$
 and a dashed line for $60^\circ$.
 (c) The effect of Mie scattering for $\log E=20$ and $\theta=0^\circ$.
 Circles are for the case that only Rayleigh scattering is
 taken into account, while triangles are for the case that Mie scattering is
 also considered.
 (d) The case of HiRes filter transmission and PMT QE for the
 showers with $\log E=20, \theta=0^\circ$. Both Rayleigh and Mie
 scatterings are taken into account. Circles are the
 same with those in (c).
}
\label{showerphoton}
\end{figure}

In Fig.\ref{showerphoton}, the ratios $R_N$ of the number of total observed
photons with the HiRes photon yields ($N_{HiRes}$)
to that with the present ones ($N_{present}$)
 are plotted for various conditions as a function of distance (horizontal 
distance to the 2 km point a.s.l. along the shower track).
$N_{HiRes}$ is smaller than $N_{present}$ by $-11$\% to $-15$\% 
depending on the distance to the shower due to Rayleigh scattering 
(Fig.\ref{showerphoton} (a)). In the case of the inclined shower, the
change of $R_N$ with distance is smaller than for the case of the
vertical shower, and the difference is between $-11$\% and $-13$\%
(Fig.\ref{showerphoton}(b)).
$R_N$ changes little if we include the effect of Mie scattering (with its weak wavelength dependence)
 (Fig.\ref{showerphoton}(c)) and 
is almost independent of the primary energy
(Fig.\ref{showerphoton}(a)). However, if
the HiRes filter transmission and the QE of the PMT are taken
into account, $R_N$ becomes
closer to unity, $-4$\% to $-7$\%, depending on the distance
(Fig.\ref{showerphoton}(d)).  This is because the transmission 
coefficient of the broad band filter used in HiRes drops
below 320nm and above 380nm where the
difference of photon yields between 
the present experiment and those assumed by 
HiRes are relatively large.

We need more a detailed analysis based on the experimental
conditions to infer the individual energy estimation of
cosmic rays, 
taking into account the wavelength dependent items (photon yields,
 scattering effects, detector performance, etc.).

\subsection{Application of the present photon yields}

In UHECR fluorescence experiments, the primary
energy is estimated from the calorimetric energy, $E_{cal}$,
with a correction for missing energy, $E_{miss}$, that carried
by neutrinos and muons, and that lost due to nuclear excitation
\cite{song00}.
$E_{cal}$ may be  determined from the experiment from
the path-length integral multiplied by the
mean ionization loss rate, $\alpha$, over the entire shower
 as
\begin{eqnarray}
E_{cal}=\alpha \int^{\infty}_0 N_{ch}(X)\mathrm{d}X \ ,
\label{e-cal}
\end{eqnarray}
where $N_{ch}(X)$ is the number of charged particles in the shower
as a function of atmospheric depth $X$ in g/cm$^2$.
Song et al. \cite{song00} showed that this technique provides
a good estimate of the primary energy of cosmic rays 
 with $\alpha$=2.19 MeV/(g/cm$^2$) by using the CORSIKA air shower
simulation program.  

Their arguments would be accepted as far as $N_{ch}(X)$ can be
determined unambiguously. In practice, however, 
it is not straightforward to convert the observed 
number of photons for each angular bin
in the camera of the detector to $N_{ch}(X)$.
This is because the photon yield depends on the 
temperature and density of air along the
trajectory of the electrons, and most particles of
low energies are not traveling parallel to the shower axis.
Alvarez-Mu\~{n}iz et al. \cite{alv03} studied the ratio of the
average track length traveled by the shower particles 
in a some depth interval and that projected onto the shower axis.
They showed that the ratio 
depends on the shower age and is 1.18 at shower maximum.
That is, $E_{cal}$ is possibly overestimated, if Eq. (\ref{e-cal})
is used without path length correction.

Instead of estimating  $N_{ch}(X)$,
Dawson  \cite{daw02} has proposed  
to use the energy deposited in the atmosphere
by the shower   per g/cm$^2$ 
of depth. 
This is determined from the number of photons, $L(X)$,
after correcting the attenuation of photons due to Rayleigh and
Mie scattering and subtracting Cherenkov
contamination.  That is, the energy deposited
by the shower in the grammage interval $\Delta X$,   
$\Delta E_{dep}(X)$, is expressed by
\begin{eqnarray}
\frac{\Delta E_{dep}(X)}{\Delta X}= \frac{1}{\Delta X}
 \sum_i \left(\frac{L_i(X)(h\nu)_i}{\Phi_i(X)}\right),
\label{e-del}
\end{eqnarray}
where $L_i(X)$ is the fraction of the flux $L(X)$ in different
wavelength bin $i$ and $(h\nu)_i$ is the photon energy of bin $i$.

 Then the calorimetric energy is estimated from
\begin{eqnarray}
E_{cal}= \int^{\infty}_0 \frac{\Delta E_{dep}(X)}{\Delta X}\mathrm{d}X \ ,
\label{e-cal-d}
\end{eqnarray}

It would be much more realistic to use Eqs. (\ref{e-del}) and
(\ref{e-cal-d}) to estimate the primary energy.
$\Phi_i(X)$ is a function of pressure (=$\rho R_{Air} T$) at 
depth $X$ 
and is expressed by Eq. (\ref{eff-p}).  Recommended values of
 $\Phi_i^{\circ}$ and $p'$ are
 listed in Table \ref{air_comp} for each wavelength.

\section{Conclusion}

Photon yields have been measured in fifteen wave bands as a function 
of pressure, for nitrogen and dry air excited by electrons of an
average energy of 0.85 MeV.  The pressure dependencies of 15 wave bands 
between 300 nm and 430 nm have been determined with our own 
measurements. The total photon yields between 300 nm and 430 nm
are 22.20$\pm$0.56 and 4.05$\pm$0.14 per meter per electron at 1013 hPa
and 20$^{\circ}$C for nitrogen and air, respectively.  If we
restrict the wave bands up to 406 nm, the corresponding values are  
 21.69$\pm$0.55 and 3.81$\pm$0.13.  The
systematic error in the measurement is 13 \%.

From the pressure dependence of photon yields, their temperature
and density dependencies of each band are determined 
for application to the energy estimation of UHECRs by the fluorescence
method.  
It would be much more realistic to use energy deposition rather 
than the number of charged particles for
each angular bin in the camera of the detector to estimate
the primary energy.  $\Phi_i^{\circ}$ and $p'$ in each wavelength 
are given
in Table \ref{air_comp} for application to estimate the energy deposition.

\section*{Acknowledgment}

We are grateful to Bruce Dawson, University of
Adelaide, for his improvement of the manuscript and his kind
advice and Fernando Arqueros, Universidad Complutense de 
Madrid, for his kind advice.
This work is supported in part by the grant-in-aid 
for scientific research No.15540290 from Japan Society
for the Promotion of Science and in part by
``Ground-based Research Announcement for
Space Utilization'' promoted by the Japan Space Forum.


\begin{thebibliography}{99}

\bibitem{nag03}
M. Nagano, K. Kobayakawa, N. Sakaki and K. Ando,
Astroparticle Physics, {\bf 20} (2003) 293. 
\bibitem{bun64}
A.N. Bunner, Ph.D. thesis (Cornell University) (1967). 
\bibitem{ueno96}
S. Ueno, Master thesis (Tokyo Institute of Technology), (1996) 
(in Japanese). 
\bibitem{kak96}
F. Kakimoto, E.C. Loh, M. Nagano, H. Okuno, M. Teshima
and S. Ueno,  Nucl. Instrum. Methods Phys. Res., \textbf{A372} 
(1996) 244.
\bibitem{hes68}
J.E. Hesser, 
J. Chemical Physics, {\bf 48} (1968) 2518. 
\bibitem{dot73}
L.W. Dotchin, E.L. Chupp and D.J. Pegg,
J. Chemical Physics, {\bf 59} (1973) 3960. 
\bibitem{rika02}
\textit{U.S. Standard Atmosphere 1976}, U.S. Government Printing Office 
(Washington D.C., 1976).
\bibitem{us66}
\textit{U.S. Standard Atmosphere Supplements,1966}, 
U.S. Government Printing Office (Washington D.C., 20402).
\bibitem{vin03}
From the slide shown by V. Rizi at the Workshop Airlight 03, December (2003).
http://www.auger.de/events/air-light-03/index.html
\bibitem{euso00}
EUSO Report on the Phase-A Study, EUSO-PI-005-1 
(ed. by L. Scarsi et al.) (2004). 
\bibitem{dav64}
G. Davidson and R. O'Neil, J. Chem. Phys. \textbf{41} (1964) 3946.
\bibitem{mit70}
K.B. Mitchell, J. Chem. Phys. {\bf 41} (1970) 1795.
\bibitem{hir70}
M.N. Hirsh, E. Poss and P.N. Eisner, Phys. Rev. A \textbf{1} (1970) 1615. 
\bibitem{ulr03}
From the slide shown by A. Ulrich at the Workshop Airlight 03, December (2003).
http://www.auger.de/events/air-light-03/index.html
\bibitem{daw02}
B. Dawson, http://www.auger.org/admin/GAP-2002-067 (2002). 
\bibitem{sak03}
N. Sakaki, M. Nagano, K. Kobayakawa and K. Ando,
Proc. Int. Cosmic Ray Conf. (Universal Academy Press, Inc., Tokyo)
	(2003) 841.
\bibitem{corsika} D. Heck, J. Knapp, J.N. Capdevielle, G. Schatz, and
              T. Thouw, Report {\bf FZKA 6019} (1998), Forschungszentrum 
              Karls\-ruhe; 
  http://www-ik.fzk.de/\~{}heck/corsika/physics\_description/corsika\_phys.html
\bibitem{corsika_upgrade} D. Heck and J. Knapp, Report {\bf FZKA 6097}
	(1998), Forschungszentrum Karls\-ruhe
\bibitem{kal97}
N.N. Kalmykov, S.S. Ostapchenko and A.I. Pavlov,
Nucl. Phys. B (Proc. Suppl.), {\bf 52B} (1997) 17. 
\bibitem{ris04}
M. Risse and D. Heck
Astroparticle Physics, {\bf 20} (2004) 661. 
\bibitem{song00}
C. Song, Z. Cao, B.R. Dawson, B.E. Fick, P. Sokolsky and X. Zhang,
Astroparticle Physics, {\bf 14} (2000) 7. 
\bibitem{alv03}
J. Alvarez-Mu\~{n}iz, E. Marqu\'{e}s, R.A. V\'{a}zquez and E. Zas,
Phys. Rev D, {\bf 67} (2003) 101303(R). 
\end{thebibliography}
\end{document}